\documentclass{article}

\usepackage{arxiv}
\usepackage[flushleft]{threeparttable}
\usepackage[utf8]{inputenc} 
\usepackage[T1]{fontenc}    
\usepackage{hyperref}       
\usepackage{url}            
\usepackage{booktabs}       
\usepackage{amsfonts}       
\usepackage{nicefrac}       
\usepackage{microtype}      
\usepackage{lipsum}
\usepackage{graphicx}
\graphicspath{ {./images/} }

\title{Magnetic Field Based Hand Tracking}

\author{
 Sizhen Bian, Kexuan Guo, Mengxi Liu, Bo Zhou, Paul Lukowicz\\
  German Research Center for Artificial Intelligence\\
  Kaiserslautern, Germany \\
  \texttt{name.surname@dfki.de} \\
}

\begin{document}
\maketitle
\begin{abstract}
Sensor-based 3D hand tracking is still challenging despite the massive exploration of different sensing modalities in the past decades. This work describes the design, implementation, and evaluation of a novel induced magnetic field-based 3D hand tracking system, aiming to address the shortcomings of existing approaches and supply an alternative solution. This system is composed of a set of transmitters for the magnetic field generation, a receiver for field strength sensing, and the Zigbee units for synchronization. In more detail, the transmitters generate the oscillating magnetic fields with a registered sequence, the receiver senses the strength of the induced magnetic field by a customized three axes coil, which is configured as the LC oscillator with the same oscillating frequency so that an induced current shows up when the receiver is located in the field of the generated magnetic field. 
Five scenarios are explored to evaluate the performance of the proposed system in hand tracking regarding the transmitters deployment: "in front of a whiteboard", "above a table", "in front of and in a shelf", "in front of the waist and chest", and "around the waist". The true-range multilateration method is used to calculate the coordinates of the hand in 3D space. Compared with the ground truth collected by a commercial ultrasound positioning system, the presented magnetic field-based system shows a robust accuracy of around ten centimeters with the transmitters deployed both off-body and on-body(in front of waist and chest), which indicates the feasibility of the proposed sensing modality in 3D hand tracking. 
\end{abstract}

\keywords{hand tracking \and magnetic filed}

\section{Introduction}

Hand positioning has been intensively explored since hands are closely associated with daily human activities. A reliable and accurate hand positioning approach could enable a series of applications\cite{bian2022the} in health care, manufactory, robotics, augmented reality, etc. 
Benefiting from the development of machine learning, the vision-based hand positioning approach has achieved impressive performance in accuracy \cite{sharp2015accurate,han2018online} and has been deployed in different scenarios like sports\cite{webster2019co}, tele-operation\cite{mizera2019evaluation}. However, the vision-based approach relies on environmental light conditions and demands powerful hardware resources for computing. Thus sensor-based approaches are also intensively explored to supply alternative hand-positioning solutions.
Table \ref{related_work} lists some of the sensor-based arm/hand/finger positioning or gesture recognition solutions from both literature and industry. The inertial measurement unit(IMU) was explored widely due to its pervasiveness in electrical devices. Wenchaun et al.\cite{wei2021real} proposed an RNN-based real-Time 3D arm tracking method by using the 6-axis IMU from a smartwatch. The elbow and wrist positions were deduced with a media error of around ten centimeters. Compared with the state-of-the-art IMU-based solutions that use either the 9-axis IMU sensor\cite{shen2016smartwatch} or the combination of a 6-axis IMU and an extra device\cite{zhou2019limbmotion}, the proposed approach improves the usability and potential for pervasiveness by not requiring a magnetometer or any extra device and achieves comparable results. However, the above mentioned IMU-based solutions require that the user not move the torso during the tracking as any torso motion could affect the tracking accuracy. 
Commercial solutions like Xsense\cite{xsens2013full} were able to track full-body motion in dynamic body states by deploying multiple IMUs on each joint of the body. Such commercial systems cost thousands of dollars and need calibration constantly to keep the tracking performance. Besides IMUs, ultrasound\cite{robotics2017marvelmind}, radar\cite{regani2020handwriting}, capacitive sensor\cite{sizhen2021capacitive, wilhelm2020ring}, strain sensor\cite{kim2020wearable}, etc, were explored to track hand/finger. The MarvelMind ultrasound positioning system could supply high tracking accuracy in 3D space. However, the solution performed unstable in a noisy environment. The high cost also limited the radar-based solution. A capacitive sensor for hand/finger tracking is a cheap and compact solution, but the sensing modality is not robust to surrounding variations. Strain sensors could also supply high accuracy in finger/hand tracking but lack generalization because of the individual differences in hand/finger size.

\begin{table*}[ht]
\centering
\footnotesize
\caption{Sensor-based Arm/Hand/Finger Tracking Systems}
\label{related_work}
\begin{tabular}{ |p{1.3cm}| p{1.5cm}| p{0.8cm}| p{1.3cm}| p{1.1cm}| p{2.3cm}| p{2.5cm}| p{3.0cm}|}
\hline
authors/ year & Sensor & Object & Sensor Position & Tracking Error & Algorithms & Advantages & limitation\\
\hline
Wenchuan et al.\cite{wei2021real} 2021 & 6-axis IMU & Arm & wrist & 7.3cm (Media Error) & RNN & accurate arm tracking with single wrist-worn IMU & body must be in static state so that IMU data are merely caused by arm motion\\
\hline

Sheng et al.\cite{shen2016smartwatch} 2016 & 9-axis IMU & Arm & wrist & 9.2cm (Media Error) & Modified hidden Markov model (HMM) & accurate real-time arm tracking with single wrist-worn IMU & compasses incorporated in wearable devices are prone to errors, especially in indoor environments; tracking on the move falters \\
\hline
Han et al.\cite{zhou2019limbmotion} 2019 & 6-axis IMU and acoustic ranging & limbs & wrist and environment & 8.9cm (Media Error)  & point clouds-based positioning & accurate real-time limb tracking & limited scenarios, extra devices for range measurement\\

\hline
Xsense \cite{xsens2013full} 2013 & IMUs & full body & all joints & 2 deg (RMS)  & sensor fusion (sensor kinematics, biomechanical model, etc.) & full body motion animation & sensors are installed in each joints; expensive\\

\hline
EL et al.\cite{el2012shoulder} 2012 & IMUs & arm & arm & 8 deg (RMS)  & unscented Kalman filter & continuously shoulder /elbow angle estimation compensating the drift error over time & only joint angle of shoulder and elbow were explored \\

\hline
MarvelMind \cite{robotics2017marvelmind} 2017 & Ultrasound & body & body and environment & 2cm (RMS)  & TOF & universal tracking system with high accuracy & unstable in noisy environment \\

\hline
Regani et al. \cite{regani2020handwriting} 2020 & 60GHz Radar & hand & environment & 2.5\% the distance to radar  & CA-CFAR (Cell Averaging-Constant False Alarm Rate) & precise, robust and calibration-free system & high cost, deployment complexity\\

\hline
Bian et al. \cite{sizhen2021capacitive} 2021 & capacitive & hand & wrist & 96.4\% recognition  & CNN & compact, running on the edge & lack of generalization and robustness \\


\hline
Regani et al. \cite{wilhelm2020ring} 2020 & capacitive & fingers & middle finger & 13.02 deg (MAE)  & LSTM & compact, low cost & low-resistive to environmental variation \\

\hline
Kim et al. \cite{kim2020wearable} 2020 & strain sensor & fingers & fingers & 1.63 deg (MAE)  & proportional summation (PS) & high accuracy for finger tracking & complexity in wearability, lack of generalization \\

\hline
\end{tabular}
\end{table*}

In this paper, we demonstrated how oscillating magnetic field positioning technology \cite{bian2020wearable, bian2020social} can be adapted into a reliable and accurate 3D hand tracking solution. 
We leveraged the fact that the magnetic field strength is attenuated at the cubic relationship with the distance. Thus by sensing the strength of a magnetic field, the distance information could be deduced. Once we know the distance information of a hand to multiple stations and the location of each station, the coordinate of the hand could be calculated. The hand tracking solution was proposed mainly considering two advantages of the modality. Firstly, low frequency oscillating magnetic field is a robust way for distance information abstracting since environmental variation could hardly deform the field\cite{bian2021induced}. Secondly, the proposed system is easy to deploy, and the cost is less than tens of dollars, enabling the magnetic field-based technique to be a generalized solution for 3D hand positioning   

Overall, we have the following two contributions:
\begin{enumerate}
\item The general concept for using the oscillating magnetic field for 3D hand tracking, and the corresponding system optimized for reliable and accurate hand tracking,
\item Evaluation of the system in both off-body and on-body scenarios gives the hand tracking result with around ten centimeters mean absolute error, showing the potential of the proposed approach for hand tracking.
\end{enumerate}

\section{Physical background, sensing prototype, and the processing approach}

\subsection{General Principle}

\begin{figure}[!t]
\begin{minipage}[t]{1.0\linewidth}
\centering
\includegraphics[width=0.45\textwidth,height=6.0cm]{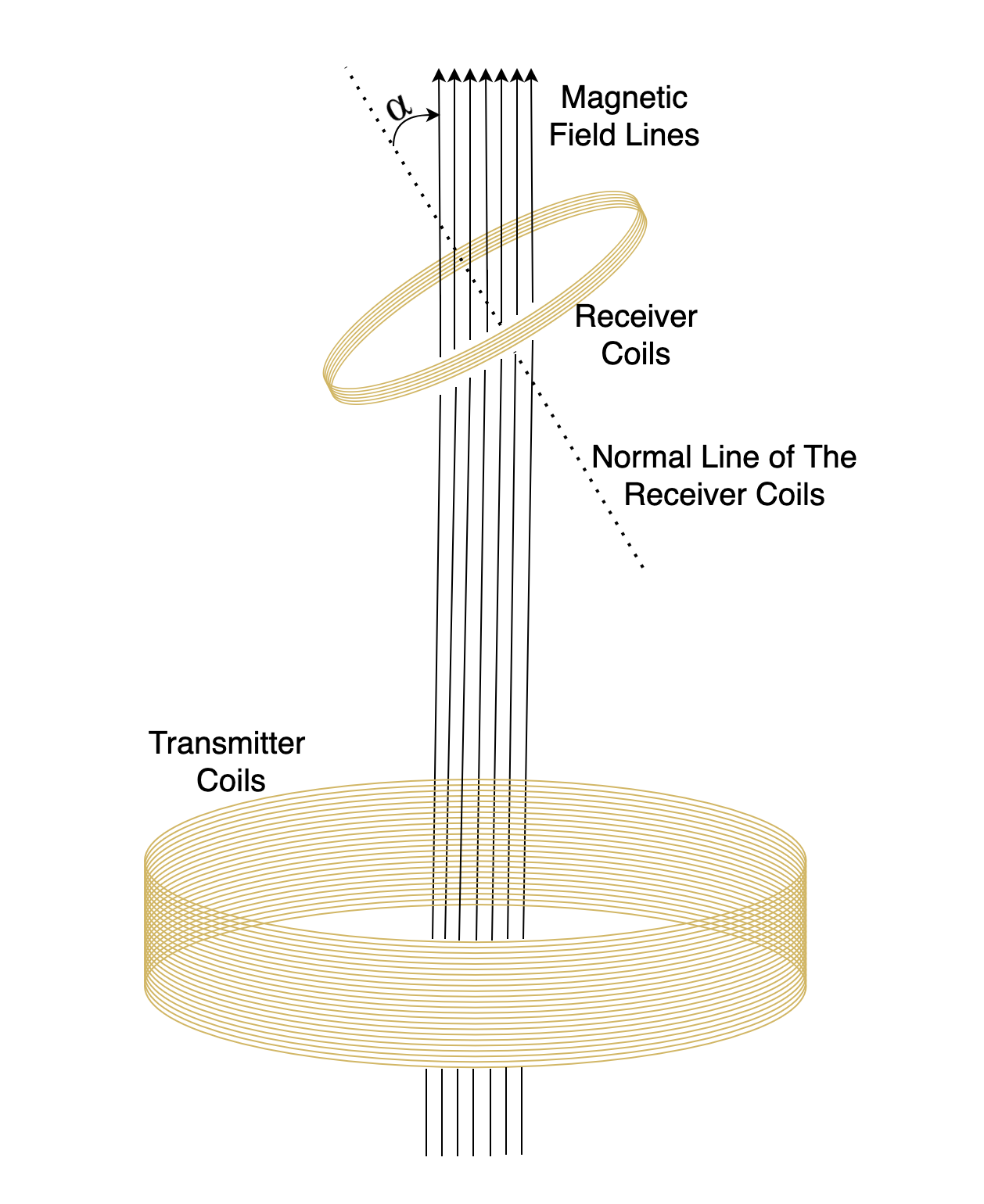}
\caption{Coils}
\label{Coil}
\end{minipage}
\end{figure}

A coil with a flowing current produces a magnetic field \(\vec{B}(\vec{r})\), where \(\vec{r}\) is the distance from the center of the coil to the measured field point. The strength of the field \(\vec{B}\) is proportional to the current \(I\) in the coil, and the direction of the field depends on \(\vec{r}\). For a resonant coil, assume that the coordinate system's origin is at the center of the coil and the $z$ axis is the normal of the coil, then the magnetic field strength along the $z$ axis is given in a simplified form by:

\begin{equation}
\vec{\mathbf{B}} = \frac{\mu_0 n a^2 I}{2\sqrt{(a^2+\vec{\mathbf{r}}^2)}^3}
\label{B_1}
\end{equation}

where $B$ is in Tesla,  $\mu_0(4\pi * 10^{-7}H/m$) is the magnetic field permeability in the vacuum, $n$ is the number of turns of the coil, $I$ is the current in the wire in Amperes, $a$ is the radius of the coil in meters, and \(\vec{r}\) is the measured point vector from the center of the coil. At a more considerable distance, the degradation of the magnetic field strength is inversely proportional to the cube of the distance. 

In an oscillating magnetic field, according to Faraday's law of induction, the peak to peak voltage accumulated at the receiver coil is proportional to the angular frequency and the amplitude of the original current, and inversely proportional to the cube of the distance from the transmitter coil to the measured position. Thus by sensing the voltage at the receiver side, distance information between the two coils can be deduced. To address the strength measurement problem when an angle between the receiver coil and the magnetic field lines exists, a tri-axis orthogonal coil at the receiver side is customized for the field strength sensing. To be noticed, the induced field is not a propagating wave, so the multi-path effects do not exist. 

\subsection{Implementation}

\begin{figure}[!t]
\begin{minipage}[t]{1.0\linewidth}
\centering
\includegraphics[width=0.55\textwidth,height=2.5cm]{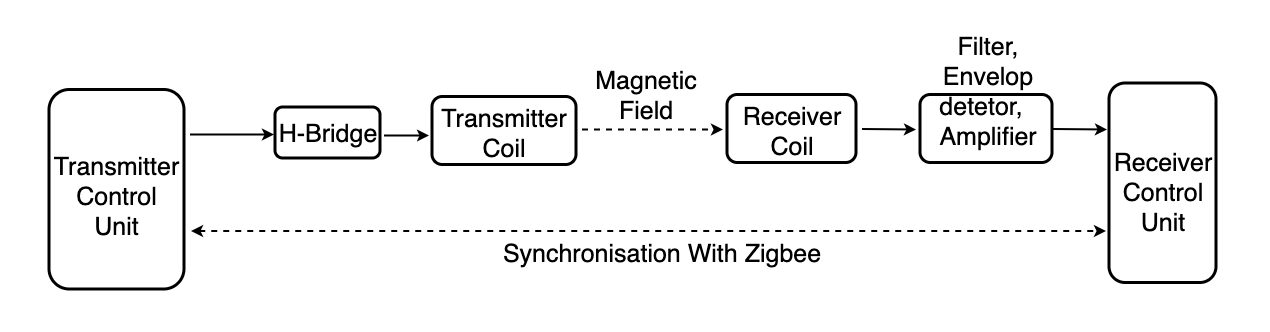}
\caption{System Structure}
\label{structure}
\end{minipage}
\end{figure}

\begin{figure}[!b]
\begin{minipage}[t]{0.5\linewidth}
\centering
\includegraphics[width=0.86\textwidth,height=5.5cm]{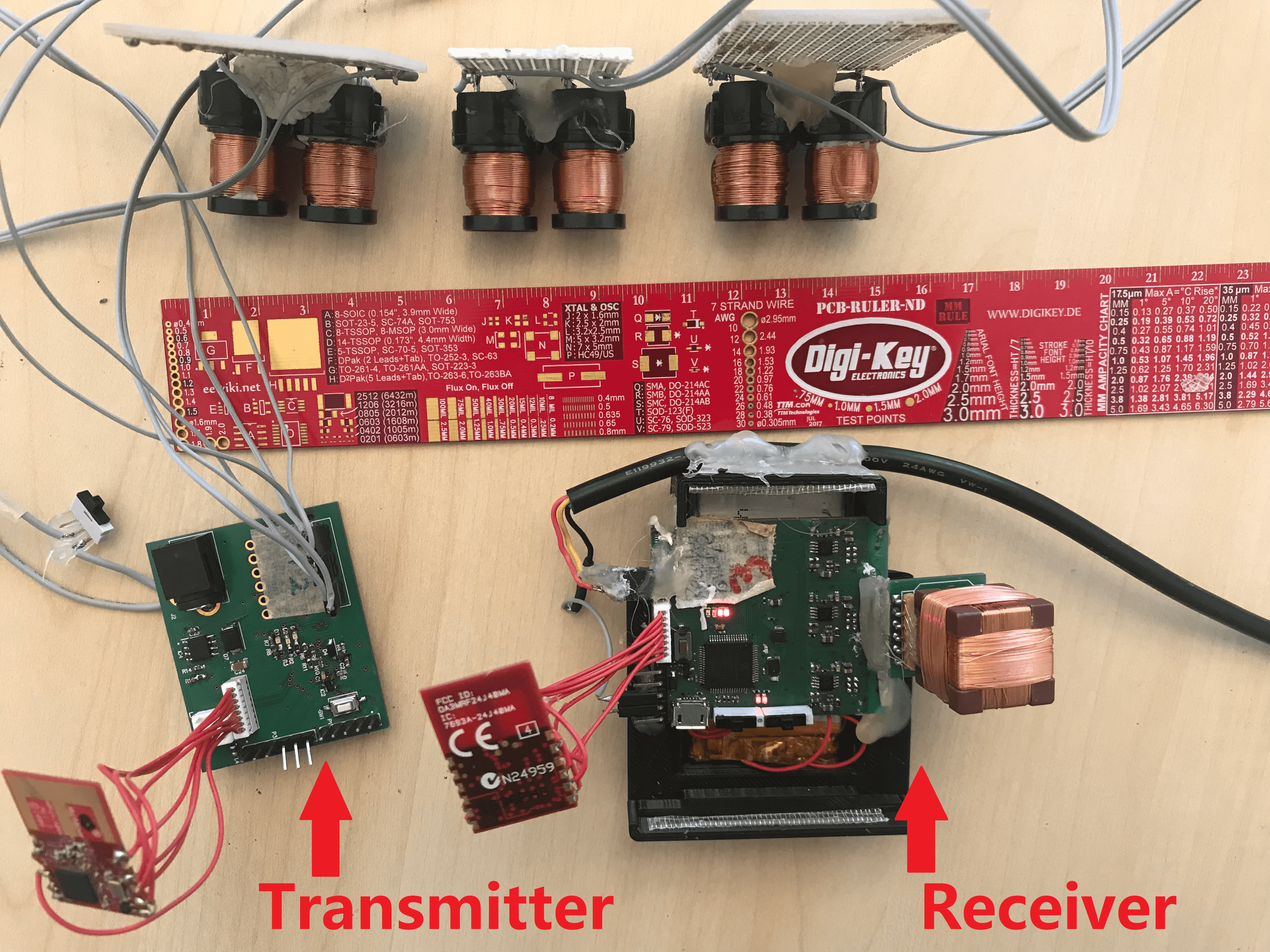}
\caption{Hardware: transmitter module, receiver module, synchronisation unit}
\label{Hardware}
\end{minipage}
\quad
\begin{minipage}[t]{0.5\linewidth}
\centering
\includegraphics[width=0.86\textwidth,height=5.5cm]{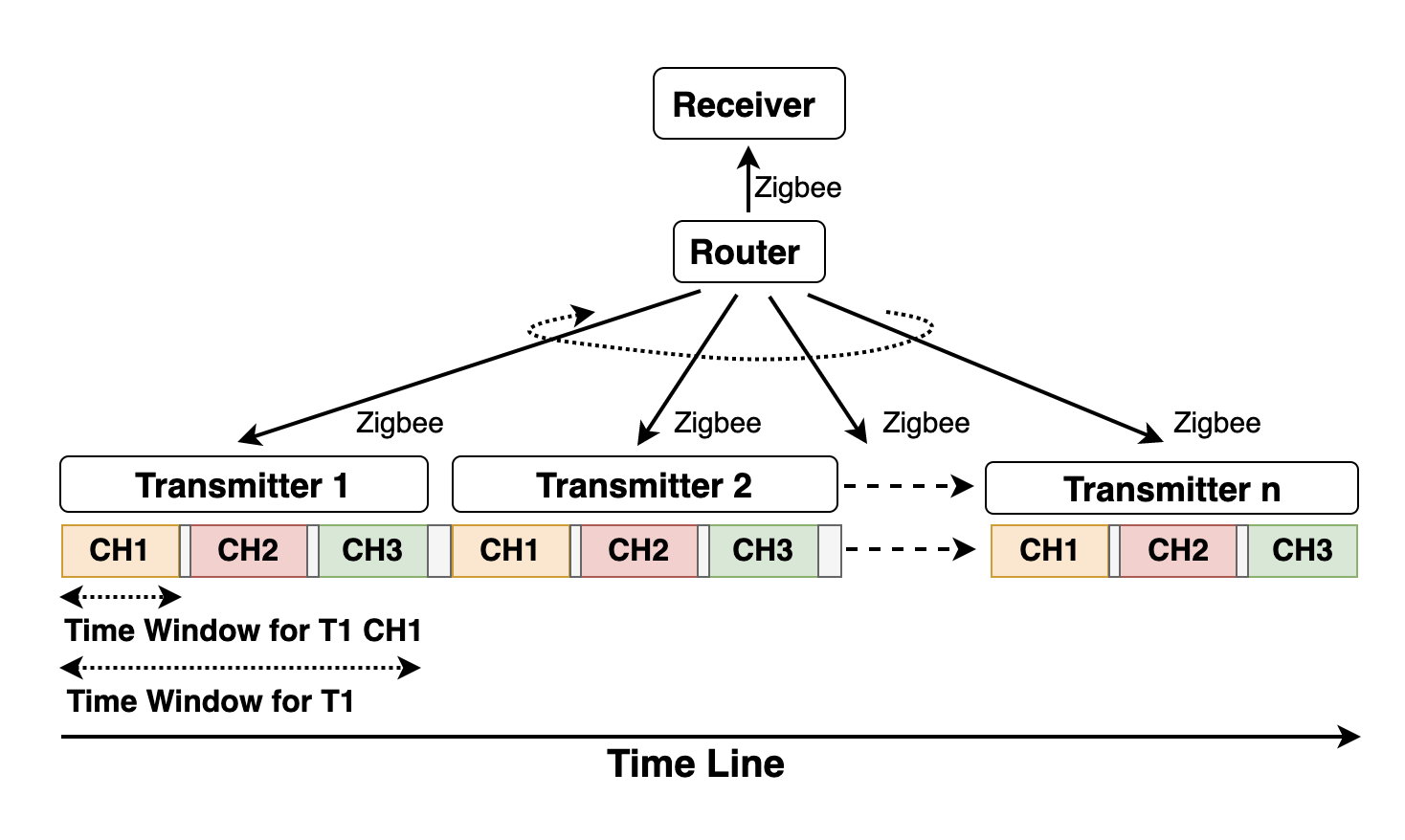}
\caption{Schedule of the being activated coils at the transmitter side}
\label{Schedule}
\end{minipage}
\end{figure}

Figure \ref{structure} and Figure \ref{Hardware} show the structure and the prototype of the proposed system. The transmitter coil, which is composed of two serially connected commercial 3.3 $mH$ inductors, generates the induced magnetic field by driving an H-bridge with a frequency of 40Khz. The chip, BD65496, with low power consumption and a compact surface-mount package from Rohm Semiconductor, is used as the H-Bridge driver to trigger the LC resonant circuit at the transmitter side. Each transmitter board could drive three transmitter coils. The receiver tri-axes coil with a value of 10 $mH$ in each axis is customized with MnZn-ferrite material in the internal space, contributing to low losses and high saturation induction. The sensed voltage on each axis of the receiver coil is then filtered and amplified by a 4-order butter worth filter and a logarithmic amplifier(AD8310), and finally sampled by a 24-bits analog-to-digital converter. The synchronization unit is accomplished by a Zigbee unit from Microchip Technology, setting up a communication channel between transmitters and receivers. For example, in a two transmitters network(including six transmitter coils and one receiver coil), one of the transmitters was initialized as a router besides the basic magnetic activation functionality. With the synchronization signal from the Zigbee module, transmitters could be activated sequentially(for example, 50 $ms$ activation time in a 70 $ms$ time window for each transmitter coil), as Figure \ref{Schedule} depicts. As a result, the receivers will have the same start time for sampling(166.7 $Hz$ sample rate for six transmitter coils). While enabling the Zigbee module, the receiver sub-system consumes 120 $mA$ current at 3.7 $V$ power input, and the transmitter sub-system consumes 84 $mA$ current at 10 $V$ power input.

\subsection{Sensing range test}

\begin{figure}[!b]
\begin{minipage}[t]{1.0\linewidth}
\centering
\includegraphics[width=0.55\textwidth,height=6.0cm]{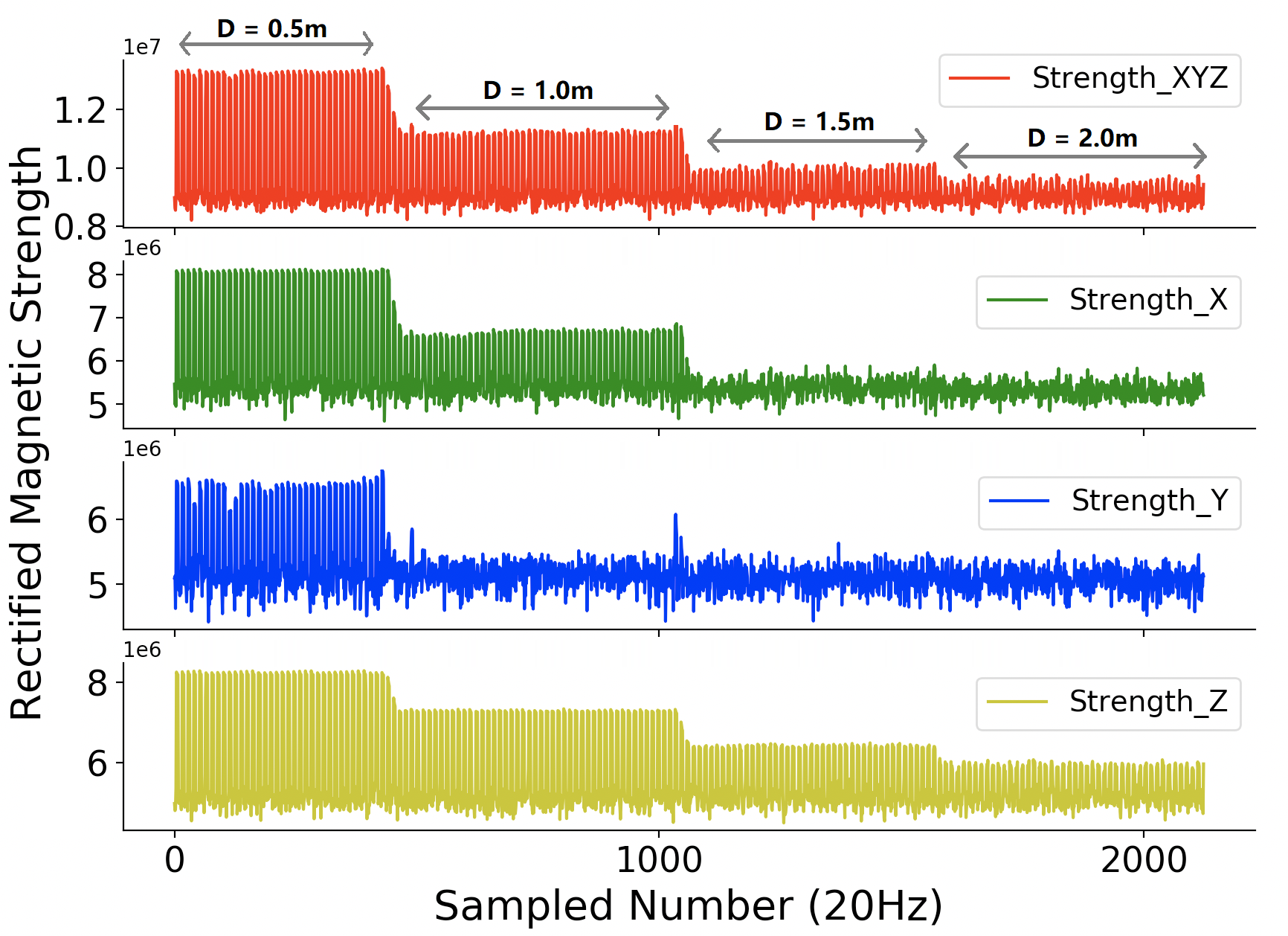}
\caption{Sensing Range Test}
\label{Range_Test}
\end{minipage}
\end{figure}

To check the sensing range of our implementation, we moved the receiver coil away from an activated transmitter coil with controlled distances. As Figure \ref{Range_Test} shows, the rectified magnetic field strength sensed by the receiver coil becomes weaker as the receiver coil moves away. Furthermore, as the top subplot shows, the sensing range of the receiver coil could reach up to two meters with current coils configuration(size, driving current, signal amplification factor, etc.). To be noticed, the rectified strength depicted in all figures in this paper is not the actual physical magnetic field strength with the unit of Tesla. Instead, it is the raw value sampled from the analog to digital converter, representing the strength in a way. The reason for using the raw sampled data to interpret the distance information is that, firstly, the real strength value is not easy to be back-deduced after the signal is processed by the analog signal processing circuit, especially being non-linearly amplified by the logarithmic amplifier. Secondly, the distance information can be deduced by a curve-fitting method based on the sampled raw data directly, thus the deduction of the real strength is not necessary.

\subsection{Data Processing}

\begin{figure}[!b]
\begin{minipage}[t]{0.5\linewidth}
\centering
\includegraphics[width=0.9\textwidth,height=5.5cm]{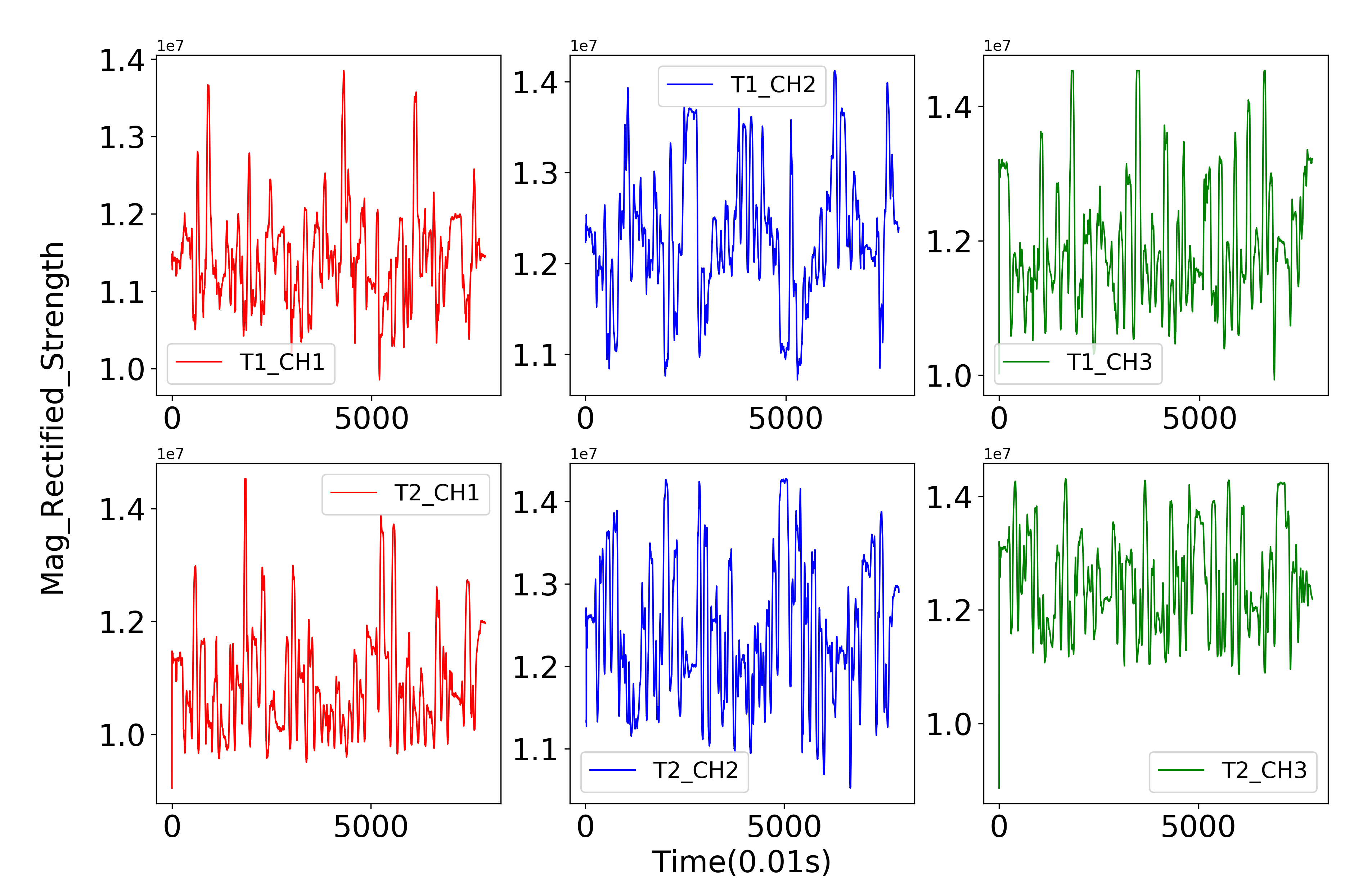}
\caption{Rectified sensed magnetic strength of the six transmitter coils}
\label{mag_strength}
\end{minipage}
\quad
\begin{minipage}[t]{0.5\linewidth}
\centering
\includegraphics[width=0.9\textwidth,height=5.5cm]{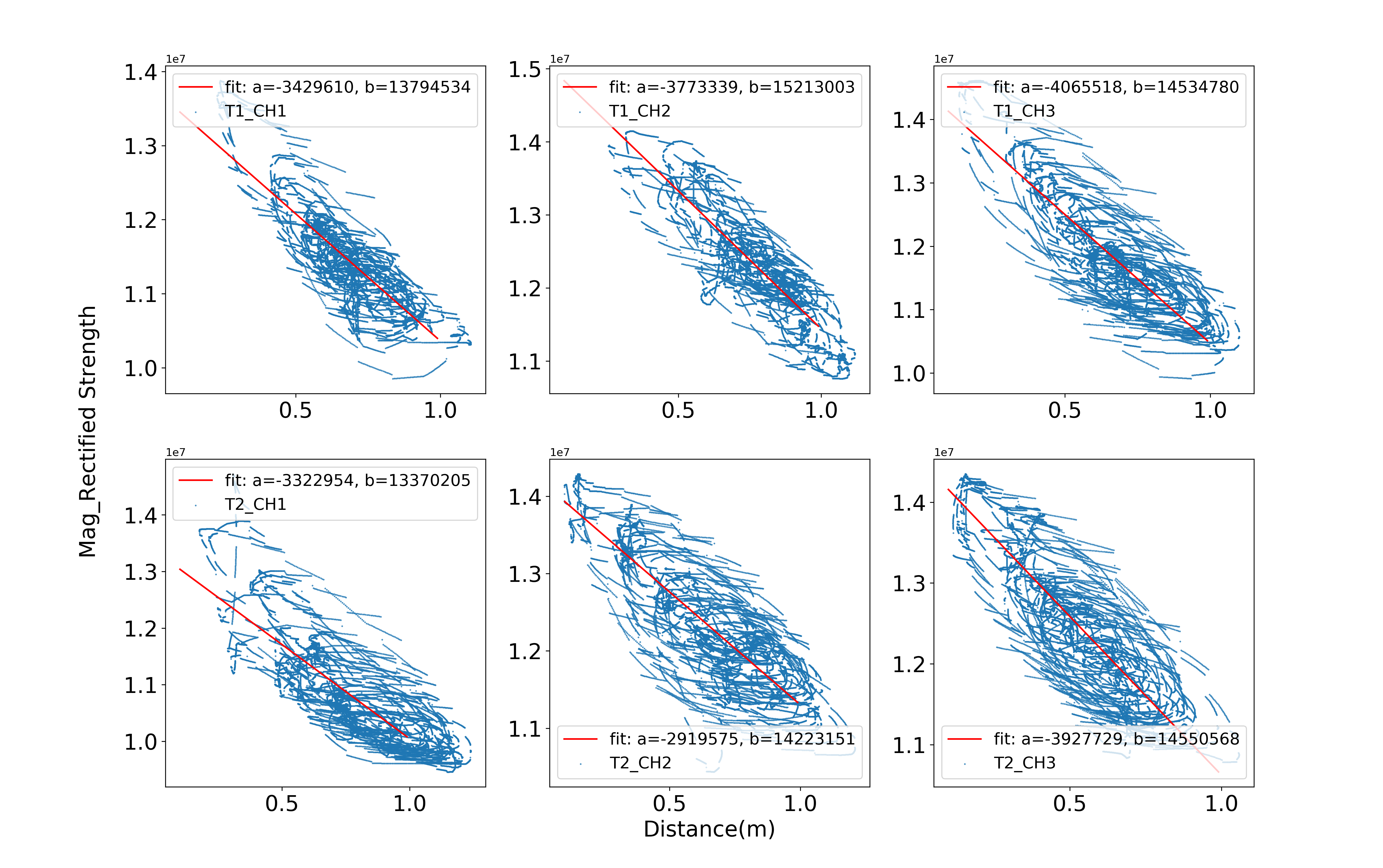}
\caption{Curve-fitting result for rectified strength vs. distance}
\label{curve_fit}
\end{minipage}
\end{figure}

\begin{figure}[!t]
\begin{minipage}[t]{1.0\linewidth}
\centering
\includegraphics[width=0.69\textwidth,height=6.0cm]{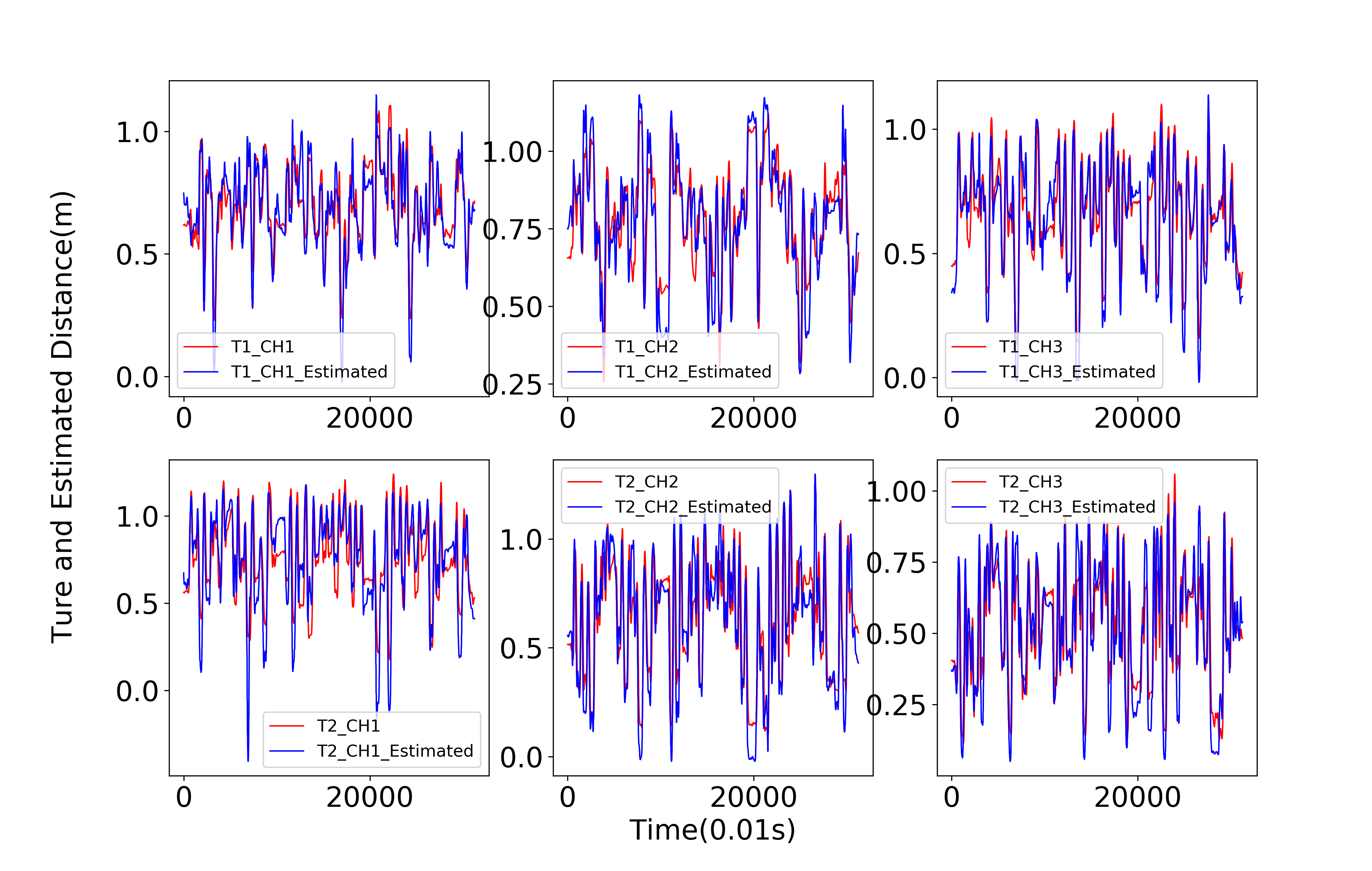}
\caption{Estimated distance(blue) vs actual distance(red) between the receiver coil and each of the transmitter coil}
\label{distance}
\end{minipage}
\end{figure}

The magnetic field of each transmitter coil was triggered with a registered sequence and specific time window. The receiver coil was held in either hand and moved in the near field of each transmitter coil to sense the magnetic field strength generated by each transmitter coil. The Zigbee unit was used to synchronize all the involved coils. Field strength data were collected either per a serial cable or a Bluetooth module to the computer with a rate of 166.7 $Hz$(when using two transmitter boards with six transmitter coils). 
The ground truth was supplied by the Marvelmind ultrasound positioning system. According to the datasheet of the commercial ultrasound system, the positioning accuracy in 3D space can reach up to 2 centimeters. However, the actual positioning error is slightly larger than the given value because of the environmental furniture and noise. Thus during the experiment, we tried to guarantee an empty and quiet space.
After the strength data was collected, as Figure \ref{mag_strength} shows, we used the curve-fitting method \cite{CurveFit} to estimate the relation between the actual distance and the rectified magnetic field strength by a linear equation($y = ax + b$) with the optimized parameters, as Figure \ref{curve_fit} depicts.  
Figure \ref{distance} shows the performance of estimated distances based on the sensed rectified magnetic field strength and the deduced linear equation of each coil by comparing it with the actual distances deduced by the ultrasound system. 

Once the estimated distances to each transmitter coil were acquired, the true-range multilateration algorithm\cite{TRM} was used to calculate the coordinates of the receiver coil. Given the coordinates of a finite number of stationaries and their distances to the moving station (derived from the sensed rectified magnetic field strength in a previous step), the true-range multilateration algorithm computes the most probable coordinates of the moving station. Even if the distances computed from each stationary do not match, the algorithm will find the coordinates that minimize the error function and returns the most optimal coordinate.
Finally, a sliding window-based smooth function was applied to the derived coordinates of the receiver coil, which also represents the coordinate of the hand.

\begin{figure}[!b]
\begin{minipage}[t]{0.5\linewidth}
\centering
\includegraphics[width=0.9\textwidth,height=5.5cm]{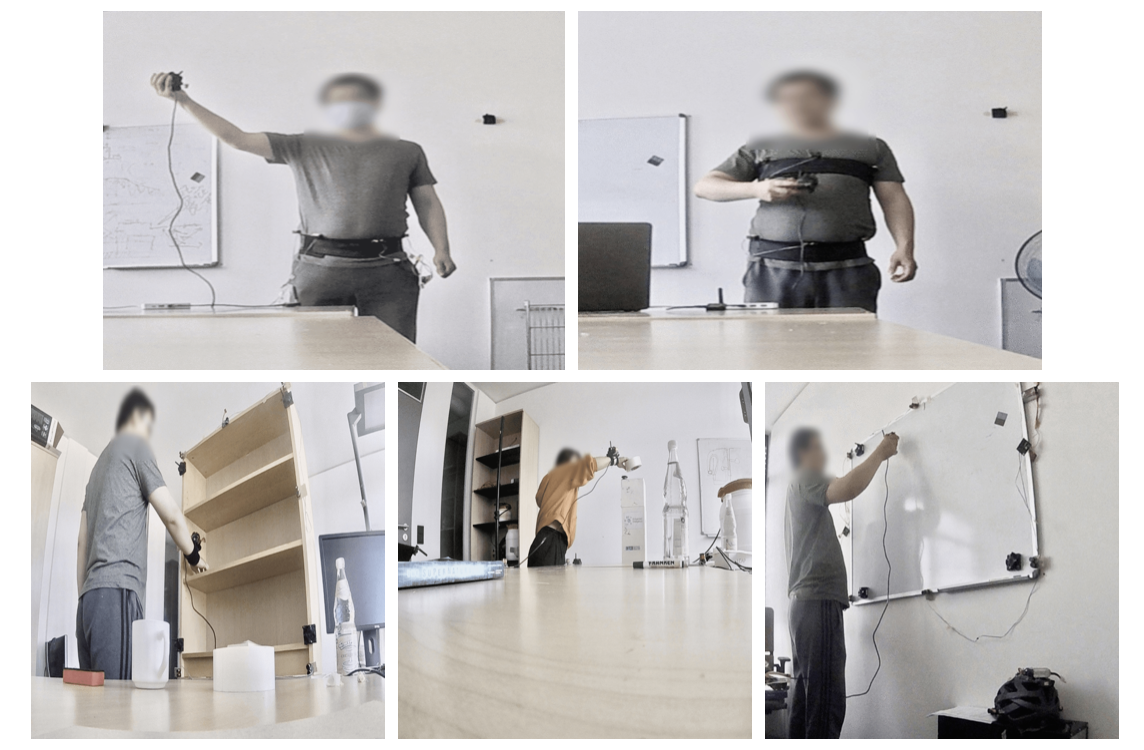}
\caption{Five case studies of hand tracking considering the deployment of the transmitter coils: "around the waist", "in front of the waist and chest","in front of and in the shelf", "above a table", and "in front of a whiteboard".}
\label{experment}
\end{minipage}
\quad
\begin{minipage}[t]{0.5\linewidth}
\centering
\includegraphics[width=0.9\textwidth,height=5.5cm]{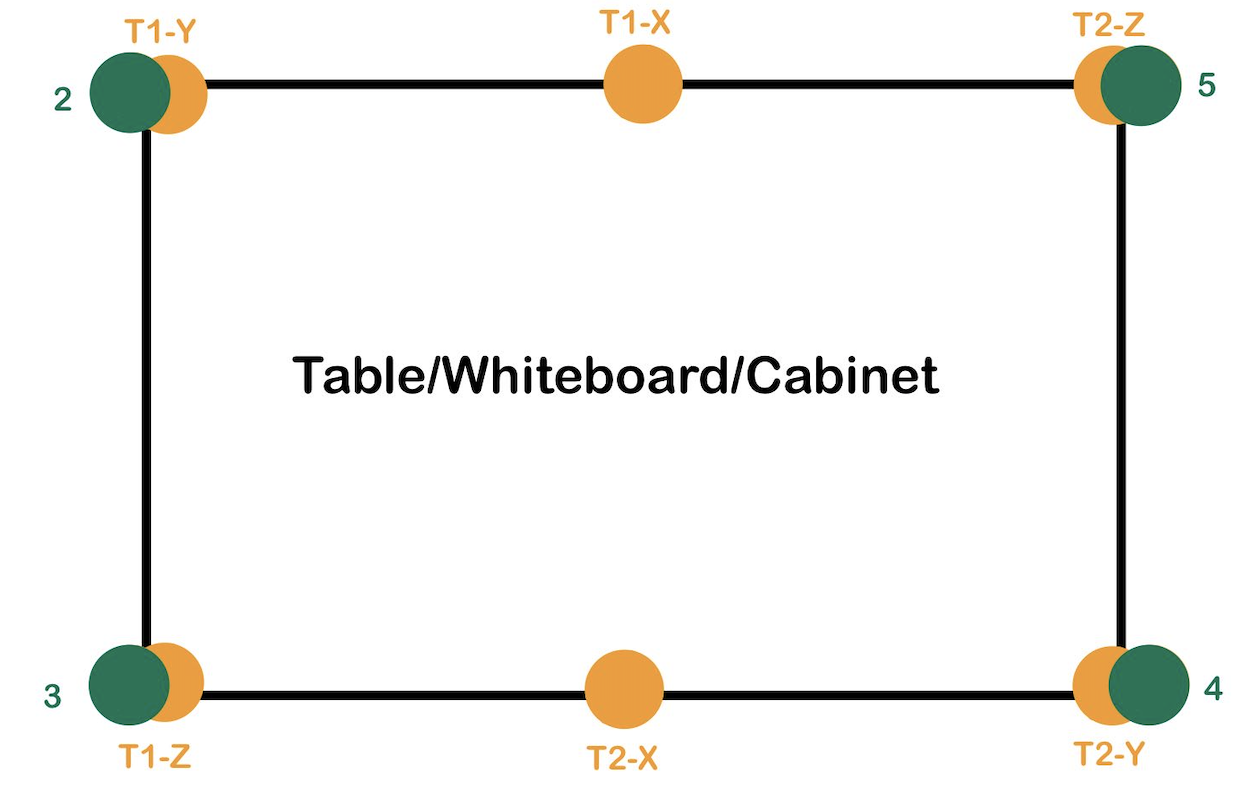}
\caption{Transmitter coils(yellow) and the ultrasound stationary beacons(green) layout in ambient}
\label{deployment}
\end{minipage}
\end{figure}

\section{Tracking cases study}

To verify the hand tracking performance with the proposed induced magnetic field-based approach, we designed five case studies as Figure \ref{experment} shows: 
\begin{enumerate}
\item Hand Tracking in front of and in a shelf
\\Move the items(pen, tape, etc.) in/off the shelf from/to a table in front of the shelf.
\item Hand Tracking above a table
\\Change the locations of different items on the table, including moving the items to a high storage box beside the table and moving back.
\item Hand Tracking in front of a whiteboard
\\Move the hand in front of a magnet whiteboard. This case study also aims to verify if a magnet object could affect the tracking performance.
\item Hand Tracking in front of the body(1)
\\Move the hand to any location that the hand could reach. Three transmitter coils were worn in front of chest and three were worn in front of waist, with a horizontal distance of around 10-18 cm, depending on the body form.   
\item Hand Tracking in front of the body(2)
\\Move the hand to any location that the hand could reach. Six transmitter coils were worn around the waist. 
\end{enumerate}

In the first three cases, two transmitter boards(six transmitter coils) were deployed around the object with the adjacent distance of 1.0-1.8 meters, as Figure \ref{deployment} shows. Four ultrasound stationary beacons were also deployed around the object for ground truth data acquisition. In the other two body-worn cases, the six transmitter coils were deployed around the body. The ultrasound system was deployed in ambient near the body. Since the coordinates of the ultrasound system was used as the reference coordinates, the body was kept in a static state so that the coordinates of the on-body transmitter coils did not variate. In a practical on-body use case,  the coordinate system from the transmitter coils will be used, thus there will be no limitation on body movement. Considering the individual body form difference, three volunteers were appealed for the body-worn case study, volunteer one(V1) with 185 cm height and 90 kg weight, volunteer two(V2) with 175 cm height and 72 kg weight, volunteer three(V3) with 160 cm height and 55 kg weight. V1 and V2 are male and V3 is female.
In all scenarios, the receiver coil was held in hand together with a mobile ultrasound beacon and moved in a random direction.

\begin{figure}[!t]
\begin{minipage}[t]{0.5\linewidth}
\centering
\includegraphics[width=0.9\textwidth,height=5.5cm]{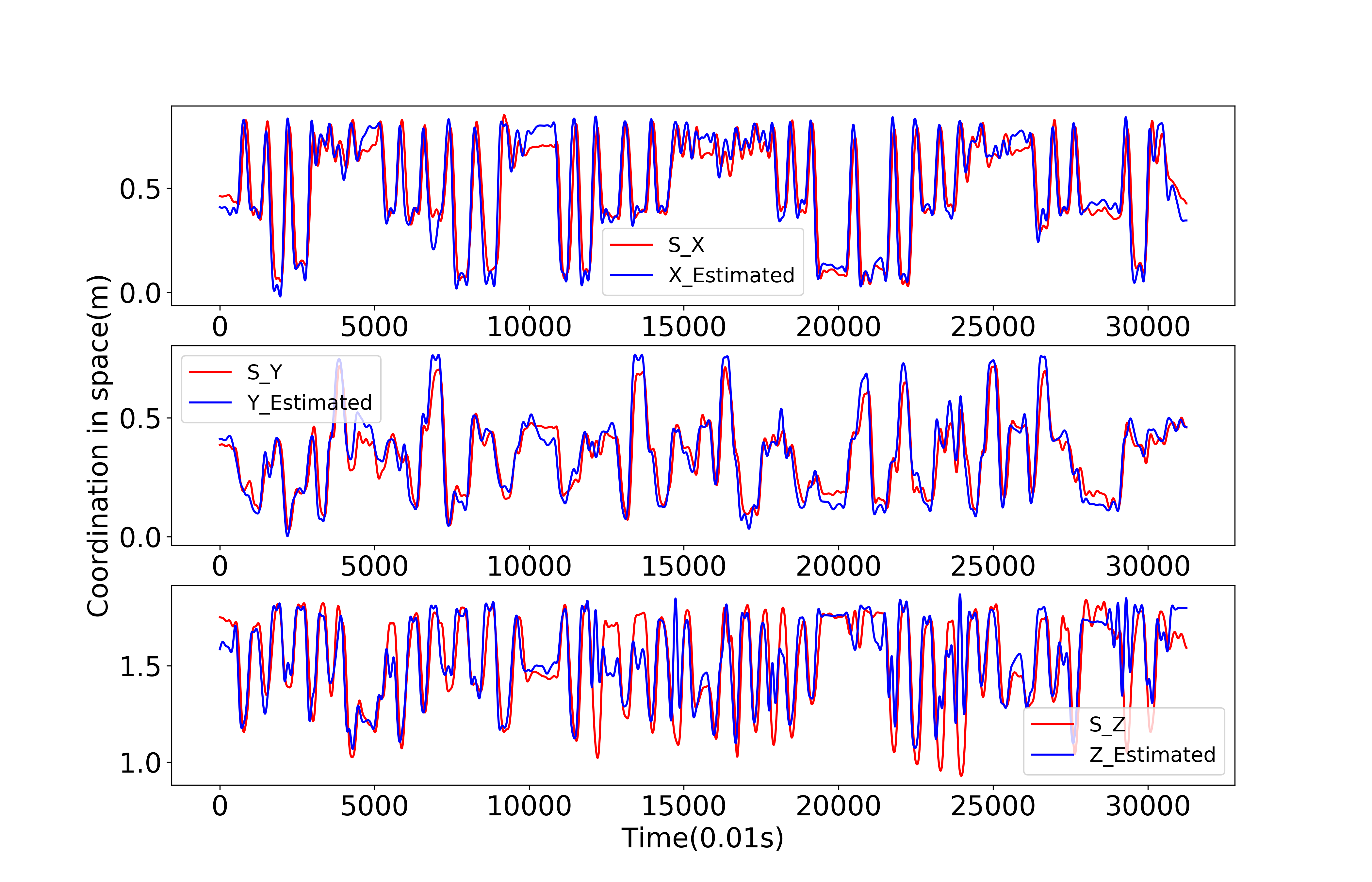}
\caption{Estimated coordinate(blue), in front of and in a shelf}
\label{coordination_cabinet}
\end{minipage}
\quad
\begin{minipage}[t]{0.5\linewidth}
\centering
\includegraphics[width=0.9\textwidth,height=5.5cm]{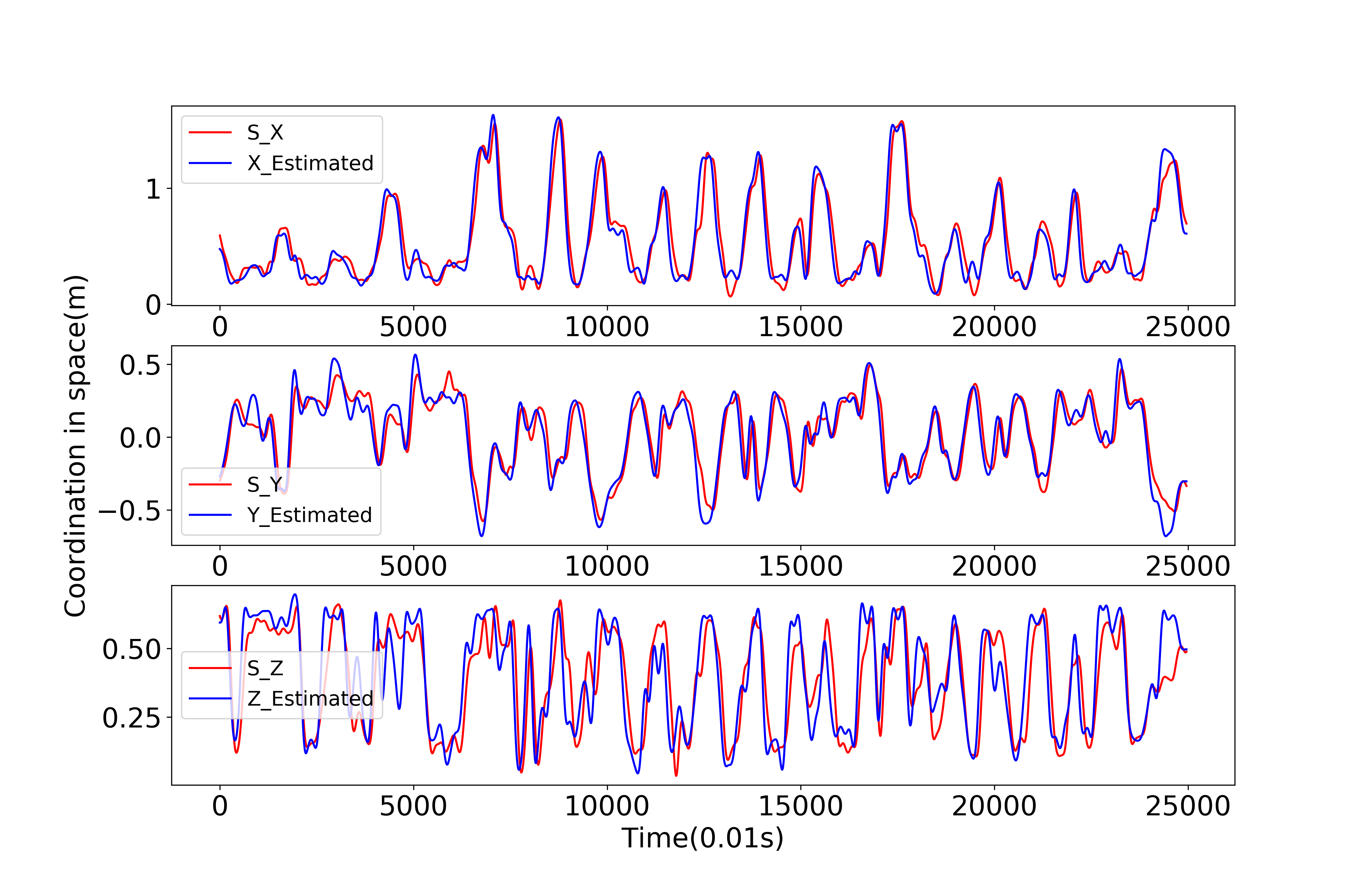}
\caption{Estimated coordinate(blue), above a table}
\label{coordination_table}
\end{minipage}
\end{figure}

\begin{figure}[!t]
\begin{minipage}[t]{0.5\linewidth}
\centering
\includegraphics[width=0.9\textwidth,height=5.5cm]{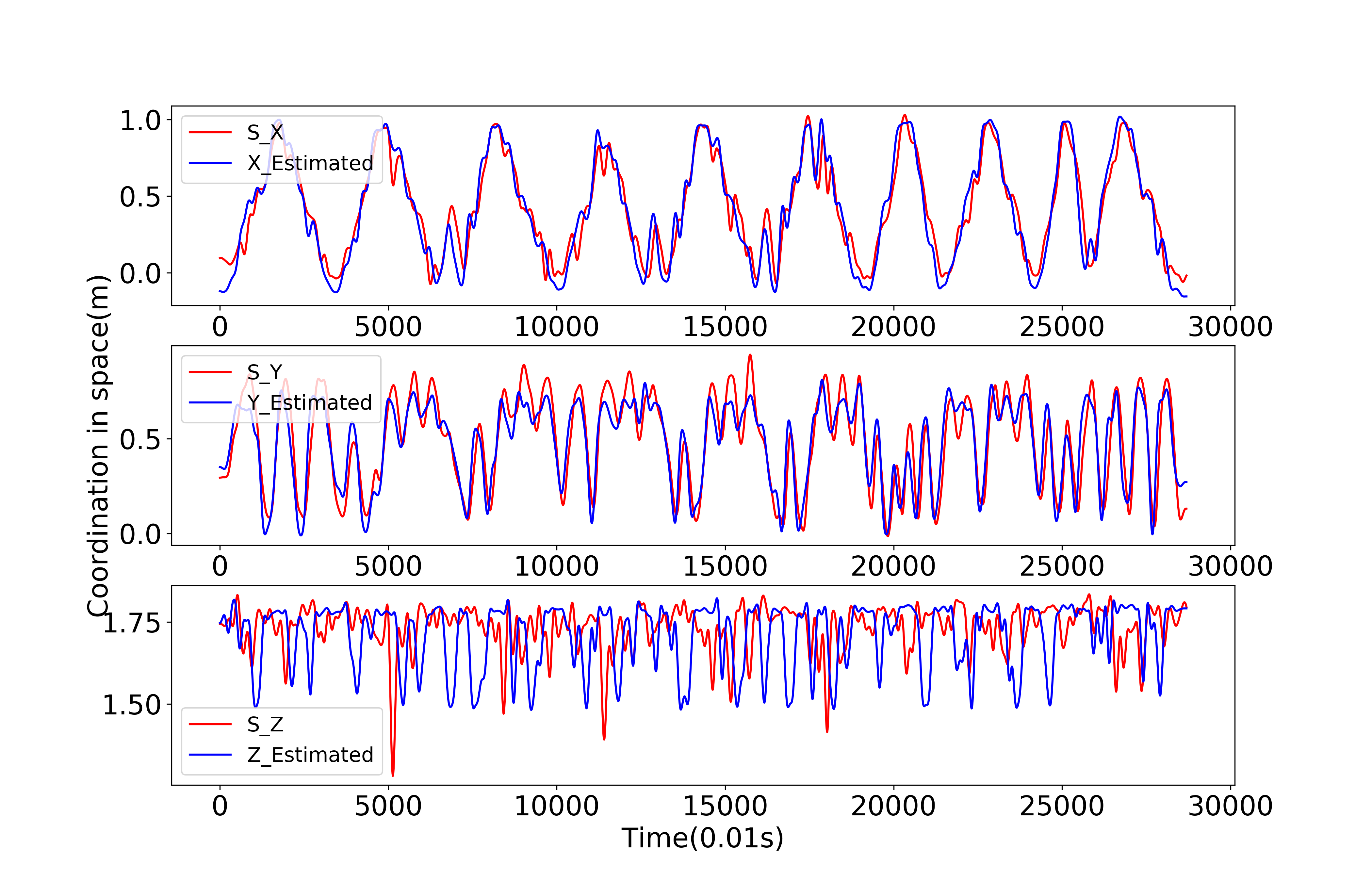}
\caption{Estimated coordinate(blue), in front of a whiteboard}
\label{coordination_whiteboard}
\end{minipage}
\quad
\begin{minipage}[t]{0.5\linewidth}
\centering
\includegraphics[width=0.9\textwidth,height=5.5cm]{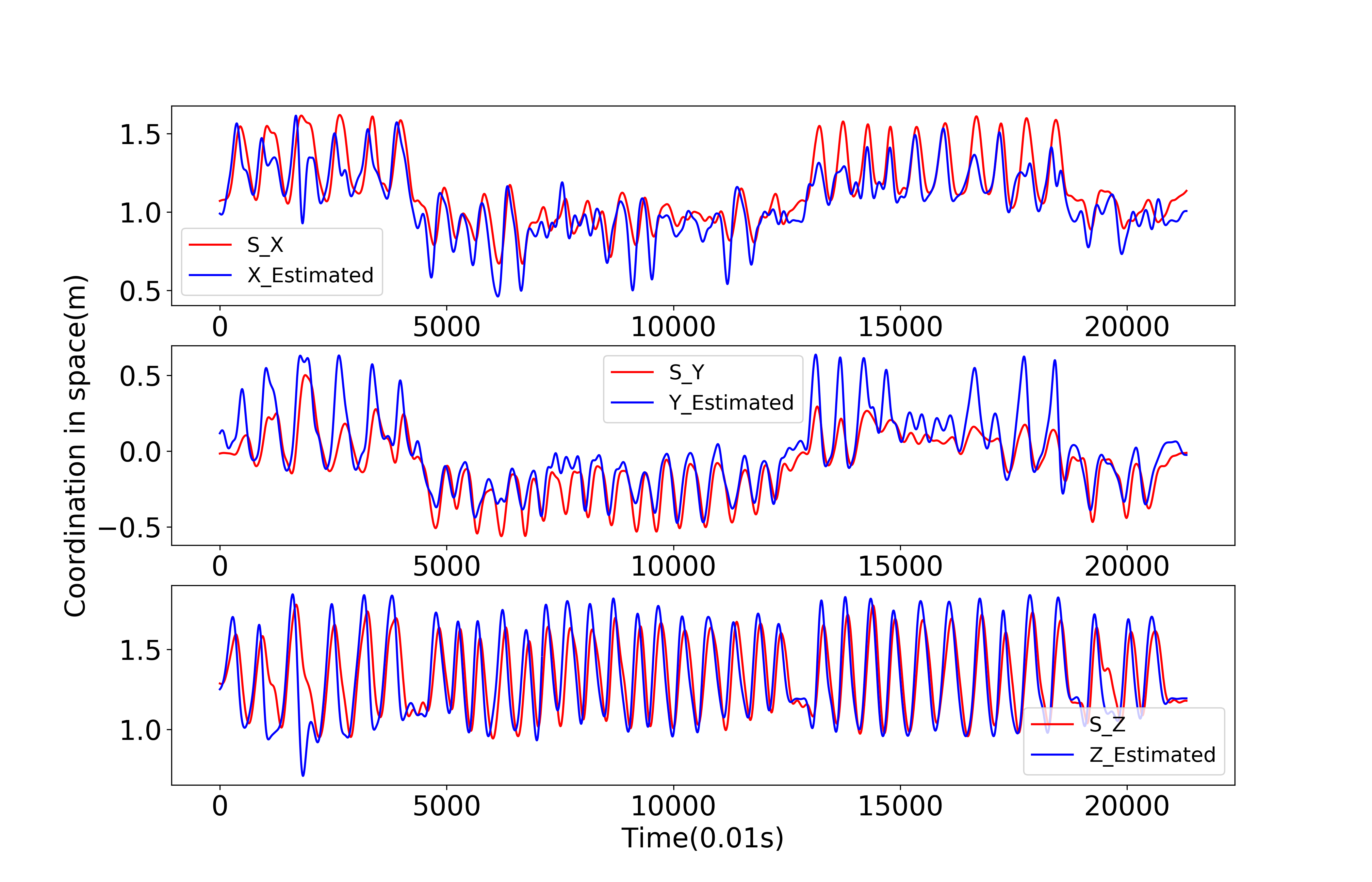}
\caption{Estimated coordinate(blue), V3, in front of the waist and chest(changed the hand during the experiment)}
\label{coordination_body_1}
\end{minipage}
\end{figure}

\begin{figure}[!t]
\begin{minipage}[t]{0.5\linewidth}
\centering
\includegraphics[width=0.9\textwidth,height=5.5cm]{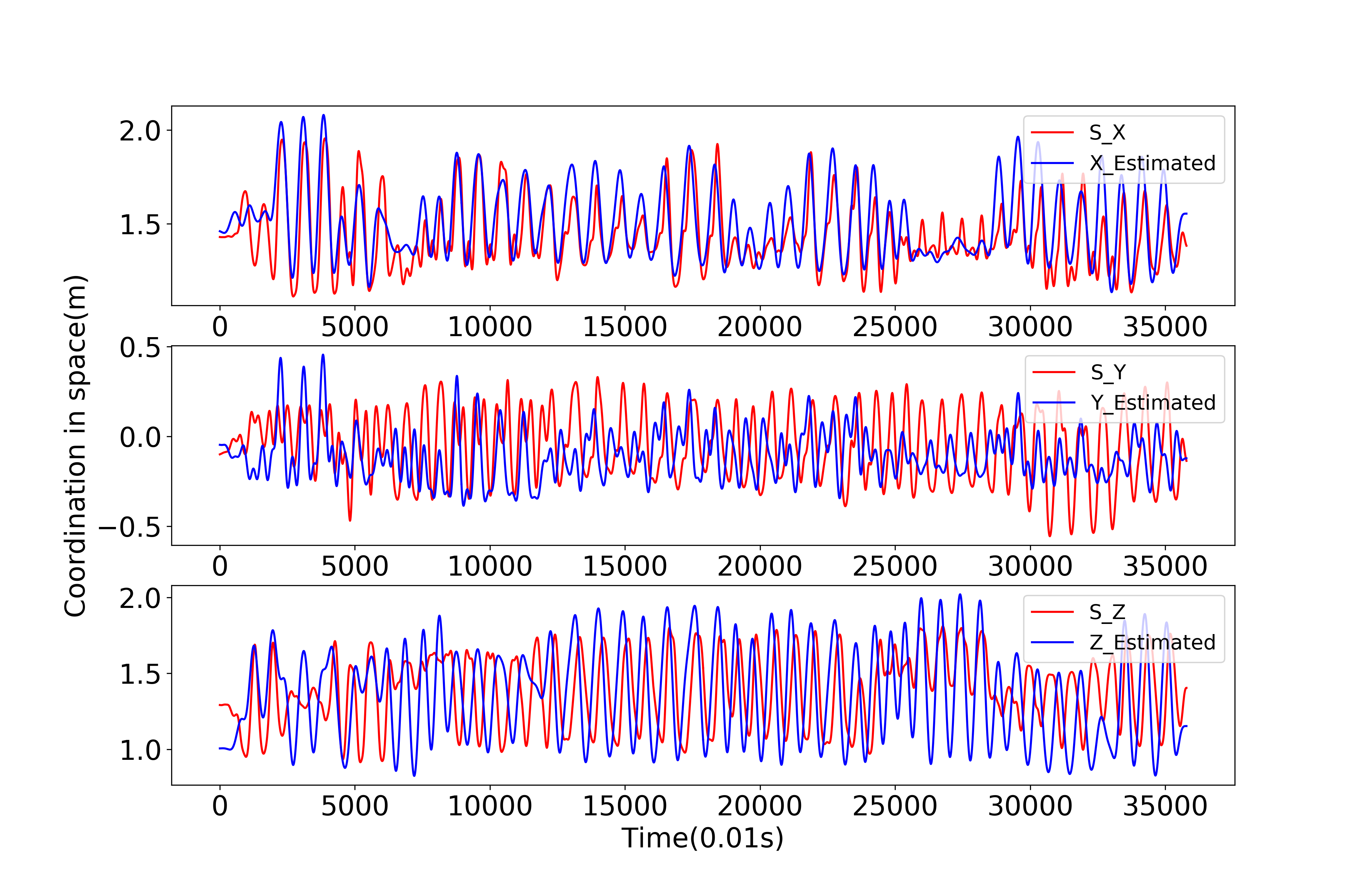}
\caption{Estimated coordinate(blue), V3, around the waist}
\label{coordination_body_2}
\end{minipage}
\quad
\begin{minipage}[t]{0.5\linewidth}
\centering
\includegraphics[width=0.9\textwidth,height=5.5cm]{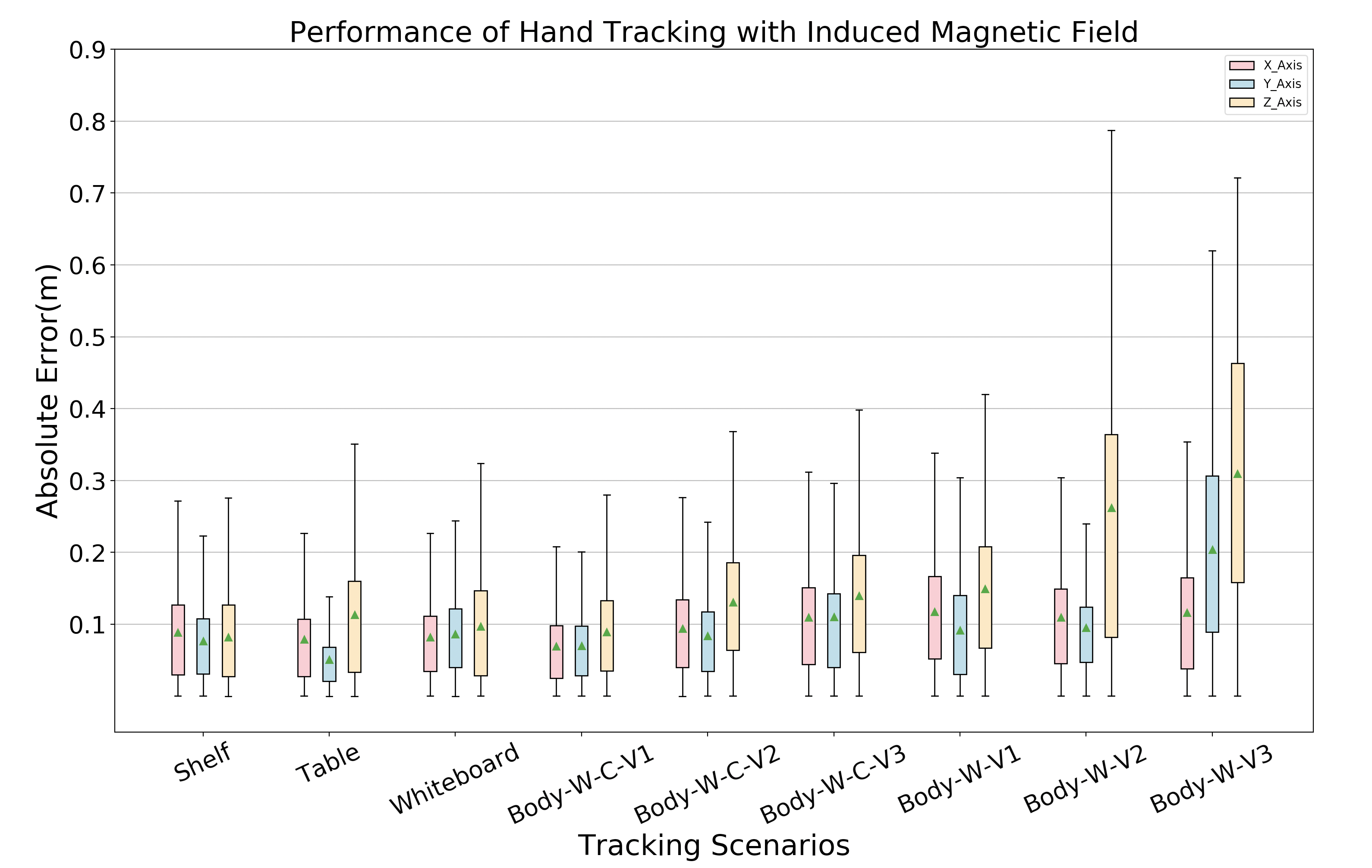}
\caption{Tracking error in different scenarios ("W\_C" indicates that the transmitter coils were deployed "in front of waist and chest" and "W" meaning "around the waist". "V1, V2, V3" are the volunteers)}
\label{track_error}
\end{minipage}
\end{figure}

Figure \ref{coordination_cabinet}, \ref{coordination_table}, \ref{coordination_whiteboard}, \ref{coordination_body_1}, \ref{coordination_body_2} show the hand tracking result(blue lines) in the five cases compared the ground truth(red lines). As can be seen, the tracking with the proposed approach performs well in all off-body case studies. The case study of the whiteboard indicates that the proposed low frequency induced magnetic field-based hand tracking approach is robust even in a magnet environment. In the two on-body studies, the "in front of waist and chest" case also shows an acceptable tracking error. However, the "around the waist" case lost the accuracy severely. Figure \ref{track_error} and Table \ref{tracking_error} give an outlook of the tracking errors from all volunteers and cases. For the "around the waist" case, only the first volunteer with the biggest body form shows acceptable 3D tracking mean absolute error of around 8-15 cm in the three axis. The second volunteer loses the tracking accuracy in Z axis, and the third volunteer loses the tracking accuracy in both Y and Z axes (over 20 cm mean absolute accuracy). The repetitions of the on-body cases show the similar errors(but no limited to Y and Z axes).

However, by looking at the estimated distance information( Figure \ref{distance}) of the "around the waist" case from V2 and V3, a severe error of distance estimation does not exist, which means that the tracking errors are related to the true-range multilateration algorithms, when interpreting the coordinate from the distance information. Considering the body form of V2 and V3, the adjacent distance of transmitter coils around the waist is short(less than 15 cm), which explains the loss of the tracking accuracy that such a coil adjacent distance is not efficient enough to estimate the coordinate by the true-range multilateration. To overcome this challenge and enable the 3D hand tracking with the proposed approach in a flexible on-body way, an alternative positioning algorithm needs to be tried, like fingerprinting, machine learning based method, etc. Another potential solution could be IMU aided sensor fusion approach. Since the distance information is reliable regardless of the transmitter coil deployment on the body, a wrist-worn IMU fused with distance information derived from a single transmitter coil could have great potential for accurate 3D hand tracking. Both solutions will be tried in our future work.


\begin{table}[!t]
\centering
\begin{threeparttable}
\caption{3D Tracking errors(MAE(Std)) in different scenarios (unit:m)}
\label{tracking_error}

\begin{tabular}{p{3.0cm} p{2.5cm}  p{2.5cm} p{2.5cm}}
\hline
Environment  & X-axis & Y-axis  & Z-axis \\\hline
Cabinet & 0.079(0.070)  & 0.051(0.041) & 0.113(0.113) \\
Table & 0.088(0.077) & 0.076(0.059) & 0.082(0.068) \\
Whiteboard & 0.082(0.062)  & 0.086(0.059) & 0.097(0.087) \\
Body(V1,W\_C) & 0.070(0.056)  & 0.070(0.055) & 0.089(0.062) \\
Body(V2,W\_C) & 0.094(0.070)  & 0.084(0.063) & 0.131(0.083) \\
Body(V3,W\_C) & 0.110(0.087)  & 0.110(0.100) & 0.139(0.101) \\
Body(V1,W) & 0.118(0.084)  & 0.091(0.074) & 0.150(0.103) \\
Body(V2,W) & 0.110(0.090)  & 0.095(0.070) & 0.262(0.252) \\
Body(V3,W) & 0.116(0.099)  & 0.204(0.131) & 0.310(0.181) \\
\hline
\end{tabular}
\end{threeparttable}
\end{table}

\section{Conclusion and future work}

This work describes how low frequency oscillating magnetic field can be used for robust and accurate 3D hand tracking in both off-body and on-body ways. We explored five hand tracking case studies with the designed prototype and validated the feasibility of the proposed approach with a tracking error of around ten centimeters. In the following exploration, we will focus on the proposed solution in a more flexible on-body way by exploring alternative positioning algorithms and the IMU-aided fusion method.

\bibliographystyle{unsrt}  
\bibliography{references}  

\end{document}